\documentclass[conference]{IEEEtran}
\ifCLASSINFOpdf
\else
\fi
\usepackage{multirow}
\usepackage{graphicx}
\usepackage{color}
\usepackage{epstopdf}
\usepackage{hyperref}
\usepackage[square,comma,sort&compress,numbers]{natbib}
\usepackage{amsthm}
\usepackage{amsmath}
\usepackage{amssymb}
\usepackage[caption=false, font=footnotesize]{subfig}
\usepackage{algpseudocode,algorithm}
\usepackage{setspace}
\newtheorem{lemma}{Lemma}

\DeclareMathSizes{10}{9}{7}{6}
\IEEEaftertitletext{\vspace{-2\baselineskip}}
\begin{document}
\title{On the Hamming Weight Distribution of Subsequences of Pseudorandom Sequences}
\author{Mahyar Shirvanimoghaddam\\
School of Electrical and Information Engineering, The University of Sydney, NSW, Australia\\
Email: mahyar.shirvanimoghaddam@sydney.edu.au}
\maketitle
\maketitle

\begin{abstract}
In this paper, we characterize the average Hamming weight distribution of subsequences of maximum-length sequences ($m$-sequences). In particular, we consider all possible $m$-sequences of dimension $k$ and find the average number of subsequences of length $n$ that have a Hamming weight $t$. To do so, we first characterize the Hamming weight distribution of the average dual code and use the MacWilliams identity to find the average Hamming weight distribution of subsequences of $m$-sequences. We further find a lower bound on the minimum Hamming weight of the subsequences and show that there always exists a primitive polynomial to generate an $m$-sequence to meet this bound. We show via simulations that when a proper primitive polynomial is chosen, subsequences of the $m$-sequence can form a good rateless code that can meet the normal-approximation benchmark.
\end{abstract}
\begin{IEEEkeywords}
Rateless codes, Gilbert-Varshamov bound, linear-feedback shift-register, finite field.
\end{IEEEkeywords}
\IEEEpeerreviewmaketitle

\section{Introduction}
A maximum-length linear sequence ($m$-sequence) is a binary sequence which satisfies the linear recurrence that is characterized by a binary primitive polynomial of degree $k$, $p(x)=\sum_{i=0}^{k}p_ix^i$ for $p_0=p_k=1$, which is referred to as the \textit{connection polynomial}. In particular, an $m$-sequence, $c_n$, can be generated by $ \sum_{i=0}^{k}p_{i}c_{n-i}=0$, where all operations are over $\mathbf{GF}(2)$. This recurrence relation generates an infinite sequence which is uniquely determined by $p(x)$. Some of the properties of these sequences have been studied for short subsequences, whose length is less than $k$. In particular, all bits in a short subsequence (of length less than $k$) are independent and the sum of these bits follows a (nearly) binomial distribution. For subsequences of lengths greater than $k$, the bits are related by the recurrence relation. Therefore, the distribution of the sum of successive bits of subsequences usually deviates from the binomial distribution \cite{Fredricsson1975}. 

In this paper, we consider all non-zero subsequences of length $n$ of an $m$-sequence generated with $p(x)$. These subsequences and the all-zero vector of length $n$ form a binary linear code of dimension $k$. We refer to this code as the \emph{Primitive Rateless} (PR) code. When $n=2^k-1$, this code is equivalent to the dual of the binary Hamming code of codeword length $2^{k}-1$ and message length $2^k-k-1$ with generator polynomial $p(x)$. Furthermore, for an arbitrary length $n\ge k$, the code is the dual of the shortened Hamming code $(n,n-k)$, where all codewords corresponding to polynomials of degree greater than or equal to $n$ are deleted from the original Hamming code \cite{Fredricsson1975}. In other words, the dual of a PR code of dimension $k$ and length $n$ is a polynomial code with generator polynomial $p(x)$. Note that $m$-sequences with connection polynomials $p(x)$ and $x^{k}p(1/x)$ are backward version of each other and hence have identical subsequence statistics \cite{Wainberg1970}. Therefore, their equivalent PR codes have the same Hamming weight distributions. 

Authors in \cite{Fredricsson1975} tried to characterize the deviation of the Hamming weight distribution of the $n$-tuples of an $m$-sequence from the truncated binomial distribution. An expression for the distribution was provided in \cite[Eq. 38]{Jordan1973}, which however depends on the primitive polynomial used to generate the $m$-sequence and is computationally complex for large $k$. Authors in \cite{Lidl1997} provided a bound for the Hamming weight of subsequences of an $m$-sequence. Finding the Hamming weight distribution of subsequences still remains a challenge and most existing approaches are computationally complex. 

In this paper, we will analyze the average Hamming weight distribution of subsequence of length $n$ of all $\phi(2^k-1)/k$ $m$-sequences of dimension $k$, where $\phi(.)$ is the Euler's totient function \cite{Lidl1997}. For this we first characterize the average Hamming weight of all dual codes (which are polynomial codes). We then use the MacWilliams Identity \cite{MacWilliams1962} to characterize the average Hamming weight of PR codes. We also derive a lower bound on the minimum Hamming weight of PR codes and show that for any $k$ and $n\ge 2k$ there exists a PR code that can meet the bound, which is identical to the Gilbert-Varshamov bound \cite{Jiang2004} for large $k$ and $n$. Moreover, when a proper primitive polynomial is chosen, the Hamming weight distribution of the PR code closely follows the truncated binomial distribution. Simulation results show that PR codes with properly chosen primitive polynomials can achieve the normal-approximation bound \cite{Polyanskiy2010}. We show that under maximum-likelihood (ML) decoding, PR codes outperforms standard Reed-Muller (RM) codes \cite{3gpprRM} for $k=3$ to $k=11$ at block lengths $n=20$ and $n=32$. We also show that using an ordered statistics decoder \cite{choi2020fast}, PR codes outperform Polar and low-density parity check (LDPC) codes with practical decoders \cite{3gpprRM,pretty_good_codes}, recently standardized for the fifth generation (5G) mobile standard.  


The rest of the paper is organized as follows. In Section II, we characterize the Hamming weight distribution of the dual of the PR code and then find the average Hamming weight distribution of PR codes. We also characterize the minimum Hamming weight of PR codes. In Section III, we study the performance of PR codes at short and moderate block lengths. Finally, Section IV concludes the paper.

\section{Hamming weight Distribution of PR codes}
Authors in \cite{Fredricsson1975} showed that subsequences of an $m$-sequence generated by using $p(x)$ as the connection polynomial, form a linear block code (we refer to it as the PR code) whose dual code is a polynomial code with generator polynomial $x^kp(1/x)$. The Hamming weight distribution of subsequences of $m$-sequences was also studied and some results were provided \cite{Fredricsson1975,Wainberg1970,Jordan1973}, which are however complex to be effectively used. Accordingly, only a little is known about the properties of so-called PR codes.

\subsection{Average Hamming weight distribution}
Our analysis here is based on the average Hamming weight distribution of subsequences of all $m$-sequences of dimension $k$. Let $\mathcal{P}_k=\{p(x): p(x)~ \mathrm{is~primitive ~and~}  \mathrm{deg}(p)=k\}$ denote the set of all binary primitive polynomials of degree $k$, where $\mathrm{deg}(p)$ denotes the maximum degree of $p(x)$. It is clear that $|\mathcal{P}_k|=\phi\left(2^k-1\right)/k$, where $\phi(.)$ is the Euler's totient function.

Let $A^{(i)}(x)$ denote the Hamming weight distribution of a PR code of length $n$ bits generated using $p_i(x)\in\mathcal{P}_k$ as the connection polynomial. It is given by
\begin{align}
    A^{(i)}(x)=\sum_{j=1}^{n} A^{(i)}_j x^j,
\end{align}
where $A^{(i)}_j$ denote the number of subsequences or codewords of Hamming weight $j$. The average Hamming weight distribution of all PR codes of dimension $k$ is then defined as
\begin{align}
    \Bar{A}(x)=\frac{1}{|\mathcal{P}_k|}\sum_{i=1}^{|\mathcal{P}_k|}A^{(i)}(x).
\end{align}

We denote the Hamming weight distribution of the dual code with generator polynomial $p_j(x)\in\mathcal{P}_k$, by $B^{(j)}(x)$. Accordingly the average Hamming weight of all dual codes is given by:
\begin{align}
    \Bar{B}(x)=\frac{1}{|\mathcal{P}_k|}\sum_{j=1}^{|\mathcal{P}_k|}B^{(j)}(x).
\end{align}
It is important to note that $m$-sequences with connection polynomials $p(x)$ and $x^{k}p(1/x)$ are backward version of each other and hence have identical subsequence statistics \cite{Wainberg1970}. Using the MacWilliams identity, we can write:
\begin{align}
   A^{(i)}_j= \frac{1}{2^{n-k}}\sum_{t=0}^{n}B^{(i)}_tK_j(t),
   \label{eq:mcwilliam}
\end{align}
where
\begin{align}
    K_j(t)=\sum_{\ell=0}^{j}(-1)^{\ell}\dbinom{t}{\ell}\dbinom{n-t}{j-\ell}
\end{align}
is the Krawtchouk polynomial \cite{BenHaim2006}, for $t$ an integer, $0\le t\le n$. Using \eqref{eq:mcwilliam}, we can easily show the following equivalence between the average Hamming weights:
\begin{align}
  \Bar{A}_j= \frac{1}{2^{n-k}}\sum_{t=0}^{n}\Bar{B}_tK_j(t),
   \label{eq:mcwilliamavg}
\end{align}
where $\Bar{A}_j=\frac{1}{|\mathcal{P}_k|}\sum_{i=1}^{|\mathcal{P}_k|}A^{(i)}_j$ and $\Bar{B}_j=\frac{1}{|\mathcal{P}_k|}\sum_{i=1}^{|\mathcal{P}_k|}B^{(i)}_j$. It is important to note that the minimum Hamming weight of any polynomial code with a primitive generator polynomial is larger than or equal to 3 \cite{Lindholm1968}, when the codeword length is sufficiently large, i.e., $B^{(i)}_1=B^{(i)}_2=0$ for any $p_i(x)\in \mathcal{P}_k$. Accordingly, one can show that a PR code of dimension $k$ and block length $n$ has the average Hamming weight equals to $\frac{n}{2}$ and the variance of the Hamming weights is $\sigma^2_n=\frac{n}{4}$.

In what follows, we first characterize $\Bar{B}(x)$ and then using \eqref{eq:mcwilliamavg} we will approximate $\Bar{A}(x)$. As stated before, $B^{(i)}(x)$ is the Hamming weight distribution of a polynomial code with generator polynomial $p_i(x)\in\mathcal{P}_k$. That is each codeword of such a polynomial code is a product of $p_i(x)$. Authors in \cite{Gupta2001,Maitra2002,Venkateswarlu2002} characterized the number of $t$-nomial (having constant term 1) multiples with degree up to $2^k-2$ of a primitive polynomial of degree $k$, denoted by $N_{k,t}$, which is given by
\begin{align}
    N_{k,t}=\frac{\dbinom{2^k-2}{t-2}-N_{k,t-1}-\frac{t-1}{t-2}\left(2^k-t+1\right)N_{k,t-2}}{t-1},
    \label{eq:tnomial}
\end{align}
where $N_{k,2}=N_{k,1}=0$. It was further shown in \cite{Gupta2001} that the distribution of $t$-nomial multiples of degree less than or equal to $2^k-2$ is very close to the distribution of all distinct ($t-1$) tuples from $1$ to $2^k-2$. Under this assumption, referred to as \emph{Random Estimate} in \cite{Gupta2001}, the probability that a randomly chosen $t$-nomial of degree at most $2^k-2$ is a multiple of a primitive polynomial is given by $N_{k,t}/{\dbinom{2^k-2}{t-1}}$ \cite{Gupta2001,Maitra2002,Venkateswarlu2002}. 

The expected number of $t$-nomial multiples having degree equals to $c$, for $c\ge\max(k,t-1)$ is then given by $N_{k,t}\dbinom{c-1}{t-2}/\dbinom{2^k-2}{t-1}$. This follows from the fact that there are exactly $\dbinom{c-1}{t-2}$ many $t$-nomials of degree $c$. 
It is also clear that when a $t$-nomial $r(x)=1+x^{i_1}+\cdots+x^{i_{t-2}}+x^{c}$ is a multiple of $p(x)$, then $x^{i}r(x)$ for $0\le i\le n-c-1$ is also a multiple of $p(x)$ and has weight $t$. There are $n-c$ of such multiples, where $\max\{k,t-1\}\le c\le n-1$. Therefore, the expected number of weight $t$ polynomials of maximum degree $n-1$, which are multiples of a primitive polynomial, is given by 
\begin{align}
    \nonumber\bar{B}_t&\approx\frac{N_{k,t}}{\dbinom{2^k-2}{t-1}}\sum_{c=\max\{k,t-1\}}^{n-1}\dbinom{c-1}{t-2}(n-c)\\
    &\overset{(a)}{\approx}\frac{1}{2^k-t}\sum_{c=\max\{k,t-1\}}^{n-1}\dbinom{c-1}{t-2}(n-c),~~3\le t\le n.
    \label{eq:dualweigth}
\end{align}
where step $(a)$ is due to $(t-1)N_{k,t}\approx\dbinom{2^k-2}{t-2}$ \cite{Gupta2001}.

\begin{figure}[t]
    \centering
    \includegraphics[width=0.95\columnwidth]{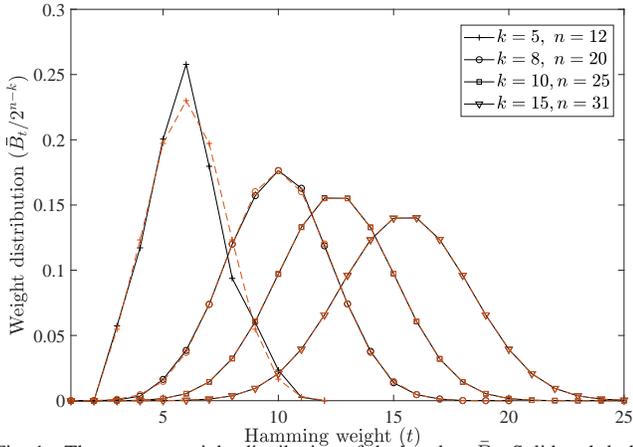}
  \vspace{-4ex}
    \caption{The average weight distribution of dual codes, $\Bar{B}_t$. Solid and dashed curves show the exact and approximate weight distributions \eqref{eq:dualweigth}, respectively.}
    \label{fig:avgweightdual}
    \vspace{-3ex}
\end{figure}

Fig. \ref{fig:avgweightdual} shows the average weight distribution of the dual of PR codes for different values of $k$ and $n$. As can be seen \eqref{eq:dualweigth} provides a tight approximation of the average Hamming weight of dual codes. There is a small gap when $k$ is small, which is mainly due to the fact that the number of primitive polynomials is also small when $k$ is small, therefore there are only a few dual codes and \emph{Random Estimate} assumption is not accurate \cite{Gupta2001}. 
\begin{table}[t]
\centering
\caption{The KLD of the exact weight distribution of the average dual code and the approximation \eqref{eq:dualweigth}.}
\label{tab1}
\scriptsize
\begin{tabular}{|c|c|c|c|}
\hline
     $k=5,~n=12$&$k=8,~n=20$ &$k=10,~n=25$&$k=15,~n=31$ \\
     \hline
     $8.3\times10^{-3}$&$6.17\times10^{-4}$&$3.09\times10^{-5}$&$1.93\times10^{-6}$\\
     \hline
\end{tabular}
\end{table}

To better characterize the approximation in \eqref{eq:dualweigth}, we use the Kullback-Leibler Divergence (KLD) to measure the distance between the exact weight distribution of the dual code and the approximation \eqref{eq:dualweigth}, which are listed in Table \ref{tab1}. As can be seen, when $k$ and $n$ go large, the approximation becomes more accurate.

We now use the MacWilliams Identity \eqref{eq:mcwilliamavg} to find the average Hamming weight of PR codes. For the simplicity of notations, we define $D_{n,t}^{(k)}:=\sum_{c=\max\{k,t-1\}}^{n-1}\dbinom{c-1}{t-2}(n-c)$. By substituting \eqref{eq:dualweigth} into \eqref{eq:mcwilliamavg}, we will have
\begin{align}
  \Bar{A}_j&\approx \frac{1}{2^{n-k}}\sum_{t=0}^{n}\frac{D_{n,t}^{(k)}}{2^k-t}K_j(t)
    \overset{(a)}{\approx} {2^{-n}}\sum_{t=0}^{n}D_{n,t}^{(k)}K_j(t),
    \label{eq:avhweight}
\end{align}
where step $(a)$ follows from $t\le n\ll2^k$. 
\begin{figure}[t]
    \centering
    \includegraphics[width=0.95\columnwidth]{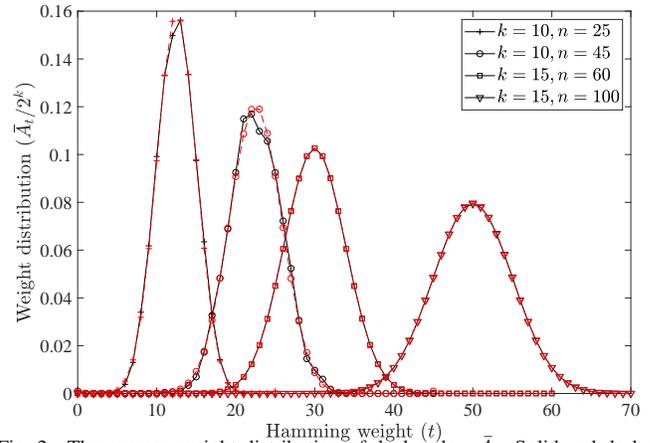}
  \vspace{-4ex}
    \caption{The average weight distribution of dual codes, $\Bar{A}_t$. Solid and dashed curves show the exact and approximate weight distributions \eqref{eq:avhweight}, respectively.}
    \label{fig:avgweight}
    \vspace{-3ex}
\end{figure}
\begin{table}[t]
\centering
\caption{The KLD of the exact weight distribution of PR codes and the approximation \eqref{eq:avhweight}}
\label{tab2}
\scriptsize
\begin{tabular}{|c|c|c|c|}
\hline
     $k=10,~n=25$&$k=10,~n=45$ &$k=15,~n=60$&$k=15,~n=100$ \\
     \hline
     $1.8\times10^{-3}$&$3.8\times10^{-3}$&$5.41\times10^{-5}$&$9.47\times10^{-5}$\\
     \hline
\end{tabular}
\end{table}
Fig. \ref{fig:avgweight} shows the average Hamming weight distribution of PR codes when $k=10$ and $k=15$. As can be seen, \eqref{eq:avhweight} provides a tight approximation for the average Hamming weight distribution. The KLD of the average Hamming weight and approximation \eqref{eq:avhweight} is shown in Table \ref{tab2}. It is important to note that the Hamming weight distribution is centered around $n/2$ with variance $n/4$. One can easily prove this for any PR code as its dual code has a minimum Hamming weight of at least 3 \cite{Lindholm1968}. We can also show that $A^{(i)}_1=A^{(i)}_2=0$ for any PR code when $n\ge2k$. We omit the proof due to limited space.

\subsection{The average minimum weight of PR code ensembles}
To characterize a bound for the minimum Hamming weight of PR codes, we first provide the following lemma.
\begin{lemma}
Let $\mathcal{C}_1$ and $\mathcal{C}_2$ denote two PR codes that are generated with primitive polynomials $p_1(x)$ and $p_2(x)$, respectively, where $p_1(x)\ne p_2(x)$ and $n\ge 2k$. Then, these codes do not have any non-zero codeword in common, i.e., $\mathcal{C}_1\bigcap \mathcal{C}_2=\{\mathbf{0}\}$.
\end{lemma} 
\begin{IEEEproof}
The lemma follows from the fact that all sequences generated by a LFSR with primitive connection polynomial $p(x)$ has linear complexity $k$. Moreover, the minimal-polynomial\footnote{The minimal polynomial of sequence $\mathbf{c}$ is the characteristic polynomial of the shortest LFSR capable of producing $\mathbf{c}$. The length of such a LFSR is referred to as the linear complexity of $\mathbf{c}$  \cite{massey1969shift}.} of the subsequence of length $n\ge 2k$ is unique \cite{massey1969shift}. Therefore, a non-zero codeword of length $n\ge 2k$ cannot be generated by two different LFSRs with different primitive connection polynomials. This completes the proof.
\end{IEEEproof} 
\begin{table*}[t]
\centering
\caption{Weight distributions of PR codes and RM codes \cite{3gpprRM}. The primitive polynomial for PR code with $k=4$ and $k=11$ are $p(x)=1+x+x^4$ and $p(x)=1+x^2+x^3+x^4+x^5+x^8+x^{11}$, respectively.}
\label{tab3}
\scriptsize
\begin{tabular}{p{0.1cm}p{0.1cm}p{1.6cm}p{13cm}}
     $N$&$k$&Code&Weight Enumerator Polynomial\\
     \hline
     20&4&RM \cite{3gpprRM}&$3x^8 + 8x^{10} + 3x^{12} + x^{20}$\\
     &&PR&$2x^9+4x^{10}+6x^{11}+3x^{12}$\\
     \hline
     32&4&RM \cite{3gpprRM}&$14x^{16} + x^{32}$\\
     &&PR&$3x^{16}+8x^{17}+4x^{18}$\\
     \hline
     20&11&RM \cite{3gpprRM}&$10x^{4} + 170x^{6} + 485x^{8} + 716x^{10} + 485x^{12}+ 170x^{14} + 10x^{16} + x^{20}$\\
     &&PR&$8 x^{4} +   29 x^{5} +   73x^{6} +   171x^{7} +249 x^{8} +  306 x^{9} +  362x^{10} +   326 x^{11} +  254  x^{12} + 161  x^{13} +  61  x^{14} +  31 x^{15} +   16x^{16} $\\
     \hline
     32&11&RM \cite{3gpprRM}&$64x^{10} + 240x^{12} + 448x^{14} + 542x^{16} + 448x^{18}+ 240x^{20} + 64x^{22} + x^{32}$\\
&&PR& $2x^{9} +    40 x^{10} +   54x^{11} +   154x^{12} +   136x^{13} +   250x^{14} +   256x^{15} +   289x^{16} +   258x^{17} +   172x^{18} +   214x^{19} +    98x^{20} +    84x^{21} +    18x^{22} + 20x^{23} + 2x^{24} $
\end{tabular}
\vspace{-5ex}
\end{table*}
\begin{figure}[t]
    \centering
    \includegraphics[width=.95\columnwidth]{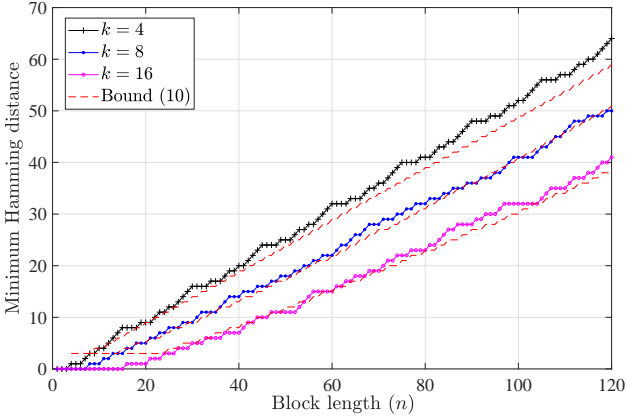}
  \vspace{-2ex}
    \caption{The minimum Hamming weight of PR codes versus the block length. The primitive polynomial for $k=4$, $k=8$, and $k=16$, are respectively, $p(x)=1+x+x^4$, $p(x)=1+x^2+x^3+x^5+x^8$, and $p(x)=1+x+x^4+x^6+x^8+x^9+x^{11}+x^{13}+x^{16}$.}    \label{fig:distancegrowth}
\end{figure}
We use Lemma 1 to derive a bound for the average minimum Hamming weight of PR code ensembles. Let us define $d_{\min}$ as follows
\begin{align}
    d_{\min}=\max_d\left\{d\left|\sum_{j=3}^{d}\Bar{A}_j\le 1 \right.\right\}.
    \label{eq:mindisbound}
\end{align}
We therefore have 
\begin{align}
    \sum_{j=3}^{d_{\min}}\sum_{i=1}^{|\mathcal{P}_k|}A^{(i)}_j\le |\mathcal{P}_k|,
    \label{eq:mindisbound2}
\end{align}
which means that the total number of length-$n$ subsequences with Hamming weight less than or equal to $d_{\min}$ of all $m$-sequences of dimension $k$ is less than $|\mathcal{P}_k|$. According to Lemma 1, the sets of non-zero subsequences of length $n$ of any two $m$-sequences of dimension $k$ are disjoint, when $n\ge 2k$, therefore, there should be at least one primitive polynomial that generates an $m$-sequence whose subsequences of length $n$ has a minimum Hamming weight larger than or equal to $d_{\min}$. We observed that when $k$ and $n$ are sufficiently large, this bounds is identical to the minimum Hamming weight obtained from the Gilbert-Varshamov bound \cite{Jiang2004} for given $k$ and $n$.

Fig. \ref{fig:distancegrowth} shows the minimum Hamming weight of PR codes at different block lengths. As can be seen, the bound \eqref{eq:mindisbound} provides a relatively accurate approximation of the minimum Hamming weight. It is important to note that for each $k$, we have used the same primitive polynomial for all block lengths $n$. One can an optimal primitive polynomials at each block length to achieve a higher minimum Hamming weight.


\section{Results and Discussion}
We first consider very short message lengths and compare PR codes with the standard Reed-Muller (RM) codes \cite{3gpprRM}. These codes have been standardized for the 5G enhanced mobile broadband (eMBB) control channel for message lengths $3\le k\le11$. Table \ref{tab3} shows the Hamming weight distribution of RM and PR codes at block lengths $20$ and $32$, when $k=4$ and $k=11$. As can be seen, the PR code with a properly chosen primitive polynomial have a lower number low-weight codewords, mainly due to its binomial-like weight distribution.   
\begin{figure}[t]
    \centering
    \includegraphics[width=0.95\columnwidth]{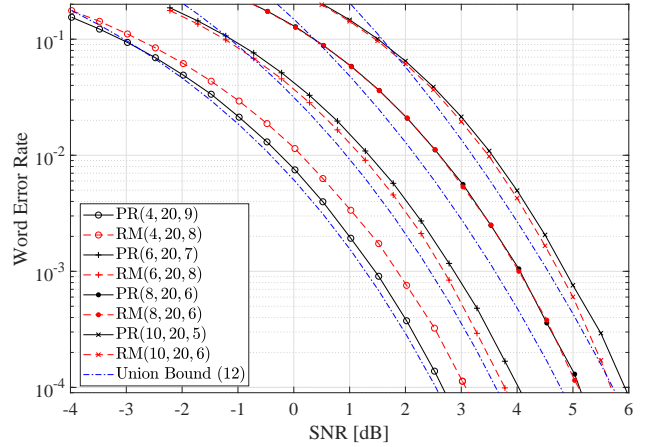}
  \vspace{-3ex}
    \caption{WER performance of RM \cite{3gpprRM} and PR codes at block length $n=20$ with ML decoding. The primitive polynomial for PR codes with $k=4, 6, 8$, and $10$ are respectively, $p(x)=1+x+x^4$, $p(x)=1+x+x^4+x^5+x^6$, $p(x)=1+x^2+x^3+x^5+x^8$, and $p(x)=1+x+x^2+x^3+x^5+x^6+x^{10}$. }
    \label{fig:lterm20}
      \vspace{-2ex}
\end{figure}

\begin{figure}[t]
    \centering
    \includegraphics[width=0.95\columnwidth]{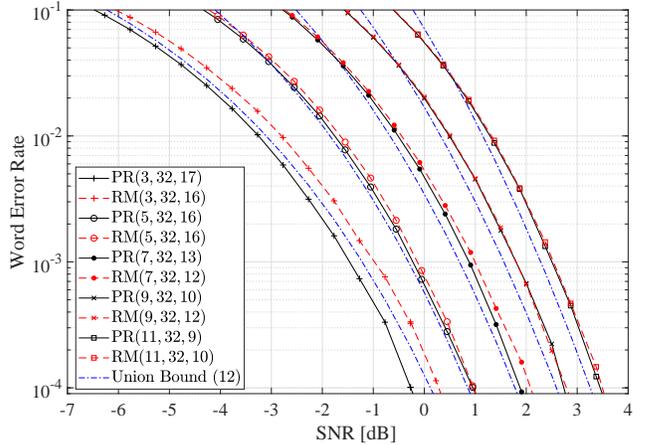}
  \vspace{-3ex}
    \caption{WER performance of RM \cite{3gpprRM} and PR codes at block length $n=32$ with ML decoding. The primitive polynomial for PR codes with $k=3, 5, 7, 9$, and $11$ are respectively, $p(x)=1+x+x^3$, $p(x)=1+x^2+x^5$, $p(x)=1+x+x^3+x^6+x^7$, $p(x)=1+x+x^3+x^4+x^9$, and $p(x)=1+x^2+x^3+x^4+x^5+x^8+x^{11}$.}
    \label{fig:lterm32}
\end{figure}

Fig. \ref{fig:lterm20} and Fig. \ref{fig:lterm32} show the word error rate (WER) of PR and RM codes at different lengths and rates under the maximum-likelihood (ML) decoding. As can be seen PR codes almost achieve the same WER performance as RM codes. In these figures, we also show the union bound (UB) which is derived using the average Hamming weight distribution \eqref{eq:avhweight} as follows:
\begin{align}
    \epsilon_{\mathrm{ub}}=\sum_{i=d_{min}}^{n}\frac{i}{n} \Bar{A}_iQ\left(\sqrt{i\gamma}\right),
    \label{eq:unionbound}
\end{align}
where $\gamma$ is the signal to noise ratio (SNR), $Q(.)$ is the standard $Q$-function, $d_{\min}$ and $\Bar{A}_i$ are obtained from \eqref{eq:mindisbound} and \eqref{eq:avhweight}, respectively. As can be seen in Fig. \ref{fig:lterm20} and Fig. \ref{fig:lterm32}, the UB \eqref{eq:unionbound} is tight when the SNR is sufficiently large. The bound is important for very short block lengths, where the normal approximation \cite{Polyanskiy2010} is loose.

\begin{figure}[t]
    \centering
    \includegraphics[width=0.95\columnwidth]{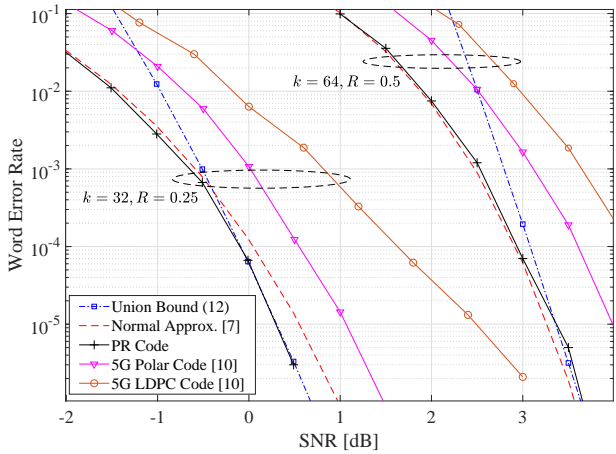}
  \vspace{-2ex}
    \caption{WER performance of PR code in comparison with 5G Polar and LDPC codes at block length $n=128$ \cite{pretty_good_codes}.}
    \label{fig:wer5gPR}
\end{figure}

\begin{figure}[t]
    \centering
    \includegraphics[width=0.95\columnwidth]{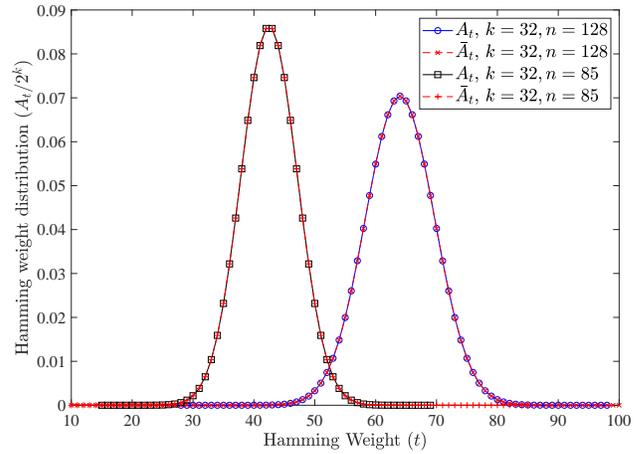}
  \vspace{-3ex}
    \caption{The Hamming weight distribution of a PR code with $k=32$. }
    \label{fig:weighdistk32}
\end{figure}

We also consider longer message lengths and compare the WER performance of PR codes with some well-known codes recently standardized for 5G eMBB. In particular, we consider the 5G Polar code for the uplink control channel with 11 bits CRC and successive cancellation list (SCL) decoding with list size 32 \cite{pretty_good_codes}. We also consider 5G low density parity check (LDPC) codes under belief propagation decoding and maximum number of iterations is set to 200 \cite{pretty_good_codes}. Results are shown in Fig. \ref{fig:wer5gPR} for the codeword length $n=128$ and message lengths $k=32$ and $k=64$. For PR codes, we have used an order-5 ordered statistics decoding (OSD) algorithm, where sufficient and necessary conditions \cite{choi2020fast} were applied to significantly reduce the decoding complexity. The primitive polynomials for the PR codes with $k=32$ and $k=64$ are $p(x)=1+x+x^2+x^5+x^7+x^8+x^9+x^{11}+x^{12}++x^{14}+x^{16}+x^{20}+x^{22}+x^{23}+x^{26}+x^{30}+x^{32}$ and $p(x)=1+x+x^2+x^3+x^4+x^6+x^7+x^8+x^9+x^{10}+x^{11}+x^{12}+x^{14}+x^{15}+x^{18}+x^{20}+x^{22}+x^{29}+x^{30}+x^{32}+x^{33}+x^{35}+x^{38}+x^{41}+x^{43}+x^{44}+x^{45}+x^{46}+x^{48}+x^{50}+x^{52}+x^{53}+x^{54}+x^{56}+x^{57}+x^{58}+x^{62}+x^{63}+x^{64}$, respectively. As can be seen in this figure, PR codes achieves a significantly lower WER than 5G Polar and LDPC codes. The PR code also closely approach the UB \eqref{eq:unionbound} at sufficiently high SNRs. Moreover, when $k=32$ and $R=0.25$, normal approximation \cite{Polyanskiy2010} is not accurate, however, UB \eqref{eq:unionbound} provides a better approximation of the minimum achievable WER at relatively high SNRs. We also show the Hamming weight distribution of the PR code with $k=32$ at different block lengths in Fig. \ref{fig:weighdistk32}. As can be seen the PR code with a properly chosen primitive polynomial has a weight distribution that can be well approximated by (\ref{eq:avhweight}), therefore the union bound in \eqref{eq:unionbound} can well approximate the WER at relatively high SNRs. 


\section{Conclusion}
In this paper, we characterized the average Hamming weight distribution of subsequences of $m$-sequences. We first found the average Hamming weight distribution of the dual code, which is a polynomial code with the generator polynomial equivalent to characteristic polynomial of the $m$-sequence. We the used the MacWilliams identity to find the average Hamming weight distribution of subsequences of $m$-sequences. We further found a lower bound on the minimum Hamming weight and showed that there always exists a primitive polynomial to generate an $m$-sequence to meet this bound. We showed via simulations that when a proper primitive polynomial is chosen, subsequences of a $m$-sequence forms a good rateless code that can meet the bound on the minimum Hamming weight.

\bibliographystyle{IEEEtran}
\footnotesize
\bibliography{ref}

\begin{thebibliography}{10}
\providecommand{\url}[1]{#1}
\csname url@samestyle\endcsname
\providecommand{\newblock}{\relax}
\providecommand{\bibinfo}[2]{#2}
\providecommand{\BIBentrySTDinterwordspacing}{\spaceskip=0pt\relax}
\providecommand{\BIBentryALTinterwordstretchfactor}{4}
\providecommand{\BIBentryALTinterwordspacing}{\spaceskip=\fontdimen2\font plus
\BIBentryALTinterwordstretchfactor\fontdimen3\font minus
  \fontdimen4\font\relax}
\providecommand{\BIBforeignlanguage}[2]{{%
\expandafter\ifx\csname l@#1\endcsname\relax
\typeout{** WARNING: IEEEtran.bst: No hyphenation pattern has been}%
\typeout{** loaded for the language `#1'. Using the pattern for}%
\typeout{** the default language instead.}%
\else
\language=\csname l@#1\endcsname
\fi
#2}}
\providecommand{\BIBdecl}{\relax}
\BIBdecl

\bibitem{Fredricsson1975}
S.~Fredricsson, ``Pseudo-randomness properties of binary shift register
  sequences (corresp.),'' \emph{IEEE Transactions on Information Theory},
  vol.~21, no.~1, pp. 115--120, 1975.

\bibitem{Wainberg1970}
S.~Wainberg and J.~Wolf, ``Subsequences of pseudorandom sequences,'' \emph{IEEE
  Transactions on Communication Technology}, vol.~18, no.~5, pp. 606--612,
  1970.

\bibitem{Jordan1973}
H.~F. Jordan and D.~C. Wood, ``On the distribution of sums of successive bits
  of shift-register sequences,'' \emph{IEEE Transactions on Computers}, vol.
  100, no.~4, pp. 400--408, 1973.

\bibitem{Lidl1997}
R.~Lidl and H.~Niederreiter, \emph{Finite fields}.\hskip 1em plus 0.5em minus
  0.4em\relax Cambridge university press, 1997, vol.~20.

\bibitem{MacWilliams1962}
F.~MacWilliams, ``Combinatorial properties of elementary {Abelian} groups ph.
  d,'' Ph.D. dissertation, thesis, Radcliffe College, 1962.

\bibitem{Jiang2004}
T.~Jiang and A.~Vardy, ``Asymptotic improvement of the {Gilbert-Varshamov}
  bound on the size of binary codes,'' \emph{IEEE Transactions on Information
  Theory}, vol.~50, no.~8, pp. 1655--1664, 2004.

\bibitem{Polyanskiy2010}
Y.~Polyanskiy, H.~V. Poor, and S.~Verdu, ``Channel coding rate in the finite
  blocklength regime,'' \emph{IEEE Transactions on Information Theory},
  vol.~56, no.~5, pp. 2307--2359, May 2010.

\bibitem{3gpprRM}
3GPP, ``{5G; NR; Multiplexing and channel coding},'' 3GPP TSG 3GPP TS 38.212
  version 15.2.0 Release 15, 7 2018.

\bibitem{choi2020fast}
C.~Choi and J.~Jeong, ``Fast soft decision decoding of linear block codes using
  partial syndrome search,'' in \emph{2020 IEEE International Symposium on
  Information Theory (ISIT)}.\hskip 1em plus 0.5em minus 0.4em\relax IEEE,
  2020, pp. 384--388.

\bibitem{pretty_good_codes}
G.~Liva and F.~Steiner, ``{pretty-good-codes.org: Online library of good
  channel codes},'' \url{http://pretty-good-codes.org}, Jan. 2021.

\bibitem{BenHaim2006}
Y.~Ben-Haim and S.~Litsyn, ``Upper bounds on the rate of {LDPC} codes as a
  function of minimum distance,'' \emph{IEEE Transactions on Information
  Theory}, vol.~52, no.~5, pp. 2092--2100, 2006.

\bibitem{Lindholm1968}
J.~Lindholm, ``An analysis of the pseudo-randomness properties of subsequences
  of long $m$-sequences,'' \emph{IEEE Transactions on Information Theory},
  vol.~14, no.~4, pp. 569--576, 1968.

\bibitem{Gupta2001}
K.~C. Gupta and S.~Maitra, ``Multiples of primitive polynomials over {GF
  (2)},'' in \emph{International Conference on Cryptology in India}.\hskip 1em
  plus 0.5em minus 0.4em\relax Springer, 2001, pp. 62--72.

\bibitem{Maitra2002}
S.~Maitra, K.~C. Gupta, and A.~Venkateswarlu, ``Multiples of primitive
  polynomials and their products over {GF(2)},'' in \emph{Proc. International
  Workshop on Selected Areas in Cryptography}.\hskip 1em plus 0.5em minus
  0.4em\relax Springer, 2002, pp. 214--231.

\bibitem{Venkateswarlu2002}
A.~Venkateswarlu and S.~Maitra, ``Further results on multiples of primitive
  polynomials and their products over {GF(2)},'' in \emph{Proc. International
  Conference of Information and Communications Security}.\hskip 1em plus 0.5em
  minus 0.4em\relax Springer, 2002, pp. 231--242.

\bibitem{massey1969shift}
J.~Massey, ``{Shift-register synthesis and BCH decoding},'' \emph{IEEE
  Transactions on Information Theory}, vol.~15, no.~1, pp. 122--127, 1969.

\end{thebibliography}

\end{document}